\documentclass{pnastwo}

\usepackage[pdftex]{graphicx}

\usepackage{amssymb,amsfonts,amsmath}
\newcommand{\comment}[1]{}

\newcommand{\ur}{URu$_2$Si$_2$}
\def \etal {{\it et al.\ }}

\contributor{Submitted to Proceedings
of the National Academy of Sciences of the United States of America}
\url{www.pnas.org/cgi/doi/10.1073/pnas.0709640104}
\copyrightyear{2008}
\issuedate{Issue Date}
\volume{Volume}
\issuenumber{Issue Number}

\begin{document}



\title{Optical spectroscopy shows that the normal state of URu$_2$Si$_2$ is an anomalous Fermi liquid}





\author{U. Nagel\affil{1}{National Institute of Chemical Physics and Biophysics, Akadeemia tee 23, 12618 Tallinn, Estonia},T. Uleksin\affil{1}{}, T. R{\~o}{\~o}m\affil{1}{},R.P.S.M.  Lobo\affil{2}{LPEM, ESPCI-ParisTech, UPMC, CNRS, 10 rue Vauquelin, 75005 Paris, France}, P. Lejay\affil{3}{Institut Ne{\'e}l, CNRS/UFJ, BP 166, 38042 Grenoble Cedex 9, France}, C.C. Homes\affil{4}{Condensed Matter Physics and Materials Science Department, Brookhaven National Laboratory, Upton, NY 11780, USA;}, J. Hall\affil{5}{Department of Physics and Astronomy, McMaster University, Hamilton, ON, L8S 4M1, Canada;}, A.W. Kinross\affil{5}{}, S. Purdy\affil{5}{}, T.J.S. Munsie\affil{5}{}, T.J. Williams\affil{5}{}, G.M. Luke\affil{5},\affil{6}{The Canadian Institute for Advanced Research, Toronto, Ontario M5G 1Z8, Canada.},  \and T. Timusk\affil{5},\affil{6}{}}
\maketitle

\contributor{Submitted to Proceedings of the National Academy of Sciences
of the United States of America}


\vspace{2cm}

\begin{article}

\begin{abstract} 
 Fermi showed that electrons, as a result of their quantum nature, form a gas of particles where the temperature and density follow the so called Fermi distribution.  In a metal, as shown by Landau, that despite their strong Coulomb interaction with each other and the positive background ions, the electrons continue to act like free quantum mechanical particles but with enhanced masses. This state of matter, the Landau-Fermi liquid, is recognized experimentally by an electrical resistivity that is proportional to the square of the absolute temperature plus a term proportional to the square of the frequency of the applied field.  Calculations show that, if electron-electron scattering dominates the resistivity in a Landau Fermi liquid,  the ratio of the two terms, $b$ has the universal value of {\em b} = 4.  We find that in the normal state of the heavy Fermion metal URu$_2$Si$_2$, instead of the Fermi liquid value of 4  the coefficient  $b$ =1 $\pm$ 0.1.  This unexpected result implies that the electrons in this material are experiencing a unique scattering process. This scattering is intrinsic and we suggest that,  the uranium $f$ electrons do not hybridize  to form a coherent Fermi liquid but instead act like a dense array of elastic impurities, interacting incoherently with the charge carriers.  This behavior is not restricted to URu$_2$Si$_2$.  Fermi liquid like states with $b \neq$ 4 have been observed in a number of disparate systems but the significance of this result has not been recognized.
\end{abstract}

\keywords{Fermi liquids | Heavy Fermions | Electron-electron scattering}





\dropcap{A}mong the heavy Fermion metals URu$_2$Si$_2$  is one of the most interesting.  It displays, in succession, no less than four different behaviors. As is shown in figure 1 where the electrical resistivity is plotted as a function of temperature, at 300 K the material is a very bad metal where the conduction electrons are incoherently scattered by localized uranium $f$ electrons. Below  $T_K \approx 75$ K, the resistivity drops and the material resembles a typical heavy Fermion metal\cite{palstra85,maple86,palstra86}.  At $T_0=17.5$ K the `hidden order' phase transition gaps a substantial portion of the Fermi surface but the nature of the order parameter is not known.    A  number of exotic models for the ordered state have been proposed\cite{chandra02,varma06,haule09,balatski09}, but there is no definitive experimental evidence to support them.   Finally, at 1.5 K URu$_2$Si$_2$  becomes an unconventional superconductor.  The electronic structure as shown by both angle resolved photoemission experiments\cite{santaner-syro09} and band structure calculations\cite{oppeneer10} is complicated with several bands crossing the Fermi surface.  To investigate the nature of the hidden order state we focus on the normal state just above the transition. This approach has been used in the high temperature superconductors where the normal state shows evidence of discrete frequency magnetic excitations that appear to play the role that phonons play in normal superconductors\cite{carbotte11}. The early optical experiments of Bonn \etal\cite{bonn88b} showed that URu$_2$Si$_2$ at 20 K, above the hidden order transition,  has an  infrared spectrum consisting of a narrow Drude peak and a strong incoherent background. The large electronic specific heat just above the transition pointed to the presence of heavy carriers with a mass $m^*=25 m_e$ \cite{maple86}. However  recent scanning tunneling microsocopy experiments contradict this model\cite{schmidt10,aynajian10}.  Schmidt \etal\cite{schmidt10} find a light band crossing the Fermi surface above 17.5 K turning into a hybridized heavy band only below the hidden order transition.  This contradicts the conventional view that mass builds up gradually below $T_K$ although there have been recent reports of some hybridization occurring in the 25 to 30 K region by Park \etal\cite{park12} and Levallois \cite{levallois11} but the reported effects are weak and perhaps not resolved by all spectroscopies. We can test the development of mass by carefully tracking the Drude weight as a function of temperature with optical spectroscopy. The Drude weight is a quantitative measure of the effective mass of the carriers. Before turning to an optical investigation of the normal state of URu$_2$Si$_2$ we will review briefly what is known from optical spectroscopy of other metallic systems at low temperature.

\begin{figure*}
\hspace*{-0.8cm}
 \includegraphics  [width=8.7cm]{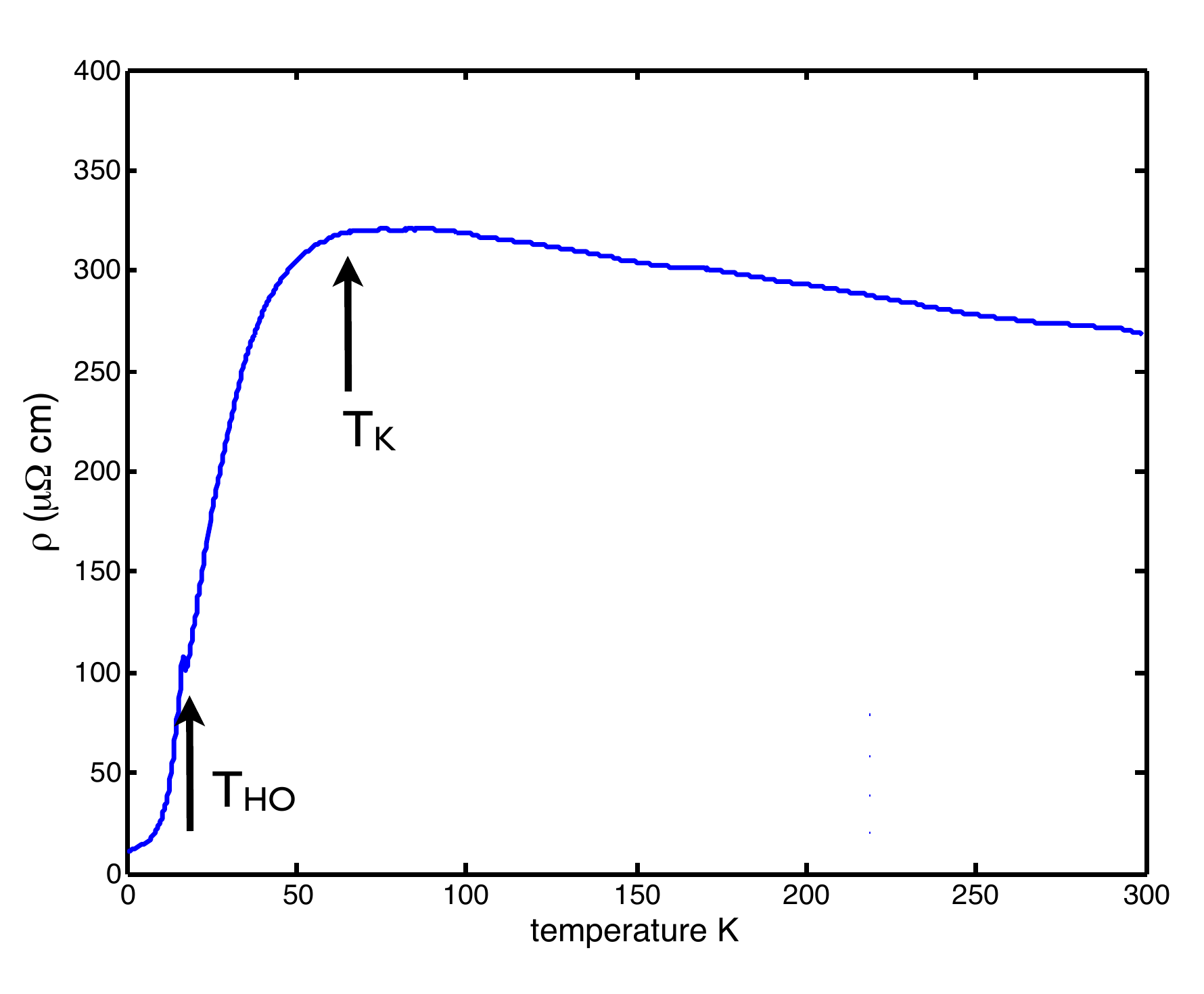}
\caption{(color online): The dc resistivity of URu$_2$Si$_2$ as a function of temperature.  Unlike ordinary metals the resistivity rises as the temperature is lowered below 300K to reach a maximum at around 75 K, referred to as the Kondo temperature, $T_K$.  Below this temperature the resistivity drops dramatically and the system acquires a Drude peak at low frequency, a defining property of a material with metallic conductivity.  This change of resistivity slope at $T_K$ is the signature of a heavy Fermion system where the conduction electrons hybridize with f electrons to form massive carriers.  In URu$_2$Si$_2$ this process in interrupted at 17.5 K by a phase transition, called the hidden order transition, where a portion of the Fermi surface is gapped.  Our aim in this work is to investigate the electrodynamics of this system just above the hidden order state.}
 \label{Fig1}
\end{figure*}

	In pure metals, at high temperature the dominant source of resistance is the electron phonon interaction giving rise to the familiar linear temperature dependence of the electrical resistance.  At low temperature the phonon contribution weakens and the resistance varies as $T^2$ where $T$ is the absolute temperature.  Gurzhi showed that under rather general conditions the resistivity of a pure metal at low temperature is given by $\rho(\omega,T) = A'(\hbar\omega^2 + 4\pi^2 (k_BT^2))$ where $\omega$ is the frequency of the field used to measure the resistivity, $T$ the absolute temperature and $A'$ a constant that varies from material to material\cite{gurzhi58}.  This formula is valid for three-dimensional systems, as long as Galilean invariance is broken by the lattice, {\it i.e.} by umklapp scattering, and for two-dimensional systems, as long as the Fermi surface is not convex and simply connected\cite{gurzhi58,millis87,rosch05,jacko09,Maslov,maslov11,pal12}, and then in the high-frequency regime, when $\omega \gg 1/\tau_{sp}(\omega,T)$ with $1/\tau_{sp}(\omega,T)$ being the singe-particle scattering rate. In the dc limit, the resistivity behaves as $\rho(T) = AT^2$, if umklapp scattering is allowed. Notice that although the coefficients $A$ and $A'$ contain different combinations of umklapp and normal scattering amplitudes, they are related as $A = 4\pi^2A'$ if umklapp scattering dominates over the normal one. We prefer to introduce a parameter $b$ which we define as $b=A/(A'\pi^2)$.  Then, if the Gurzhi resistivity formula is valid, $b=4$.  A source of confusion in the literature is the formula for the single particle scattering rate $1/\tau^{sp}$ within Fermi liquid theory $1/\tau^{sp}(\omega,T)=A'((\hbar\omega)^2+\pi^2(k_BT)^2)$ that is sometimes used to describe the resistivity.  This is wrong and to be general we will use the parameter $b$ as a quantity that is measured by comparing, in the same energy range $\hbar\omega \approx k_BT$, the frequency and temperature terms in Gurzhi's formula.  While the focus of this paper is an accurate determination of $b$ in the normal state above the hidden order transition of  URu$_2$Si$_2$, it is useful to look at previous work where the quantity $b$ can be extracted from the measured optical resistivity $\rho(\omega,T)$ and, in some cases the dc resistivity $\rho(T)$.  These are challenging experiments since Fermi liquid scattering, in most metals, is a low temperature phenomenon and therefore to stay in the energy range where the temperature dependence of the resistivity is examined, the optical measurements have to be carried out in the very far infrared, an experimentally difficult region.  Nevertheless a search of the literature turns up several examples.
	
\begin{figure*}
\vspace*{-0.5 cm}
\hspace*{-0.8cm}
 \includegraphics  [width=8.7cm]{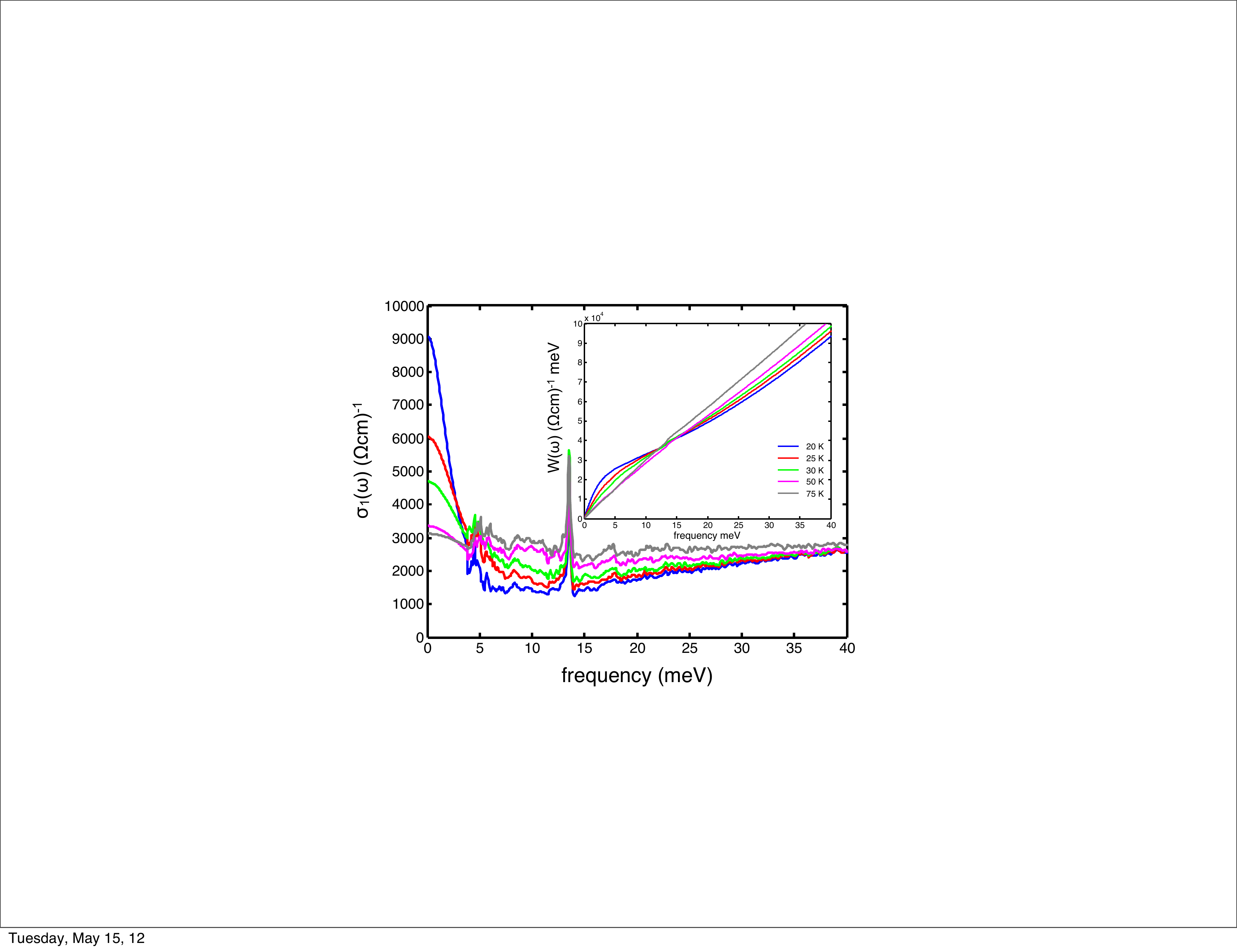}
 \vspace*{-2.0 cm}%
\caption{(color online): The optical conductivity as a function of photon energy in the heavy Fermion state.  Below 4 meV the conductivity has been fitted to a Drude peak whose amplitude agrees with the dc resistivity. The inset shows the integrated spectral weight up to a frequency $\omega$. Below 10 meV the Drude weight dominates. The total Drude weight is temperature independent since all the curves join at 15 meV above the Drude cutoff.  The spectral weight in the hybridization gap region, 5 to 40 meV, is lost to higher frequencies and the cumulated spectral weight drops at 40 meV as the temperature is lowered.}
 \label{Fig2}
 \vspace*{2.0cm}
\end{figure*}

The first report of a discrepancy of the ratio of the amplitudes of the frequency and temperature terms in a Fermi liquid was a report by Sulewski \etal\cite{sulewski88}  on the infrared properties of the heavy Fermion material UPt$_3$. Instead of the expected value of $b=4$ they reported and experimental upper limit of $b=1$. Since then a number of studies have presented both $T^2$ and $\omega^2$ dependencies of the optical scattering on the same material\cite{katsufuji99,basov02,yang06,Reedyk12} .  A summary of these is given in Table 1.  In some cases the authors have not calculated the ratio $A/A'$ in which case we have made an estimate from the published curves. We have also tabulated the approximate maximum temperatures and frequencies where the quadratic dependence is observed.  It is important that these overlap to some extent. The overall conclusion one can draw from this table is that in no case has the expected canonical Fermi liquid behavior with $b=4$ been observed experimentally. Additional examples of non Fermi liquid behavior are given in a review of Dressel\cite{dressel11}.

\begin{table*}
\vspace*{1.0cm}
\begin{tabular}[t]{lrrccc}
Table 1. Summary of experimental measurements of the ratio $b$ of temperature and frequency \\terms for some Fermi liquids. $T_{max}$ and $\omega_{max}$ indicate the upper limit of the measured \\quadratic behavior of $\rho(T)$ and $\rho(\omega)$ respectively. \\& & & \\
Material	& $T_{max}$	&$\omega_{max}$ &$b$&ref.\\
	&meV&meV& &\\ \hline\
UPt$_3$	&1&1&$<1$&\cite{sulewski88}\\
CePd$_3$& & & 1.3&\cite{sulewski88}\\
Ce$_{0.95}$Ca$_{0.05}$TiO$_{3.04}$&25&100&1.72&\cite{katsufuji99}\\
Cr	&28&370&2.5&\cite{basov02}\\
Nd$_{0.95}$TiO$_4$&24&50&1.1&\cite{yang06}\\
URu$_2$Si$_2$	&2&10&1.0&present work\\

\end{tabular}
\caption{Summary of experimental measurements of the ratio $b$ of temperature and frequency terms for some Fermi liquids. $T_{max}$ and $\omega_{max}$ indicate the upper limit of the measured quadratic behavior of $\rho(T)$ and $\rho(\omega)$ respectively. }

\end{table*}

\section{Results}
Figure 2 shows the optical conductivity between 20 and 75 K, the region where coherence develops as shown by the appearance of a Drude peak below 15 meV which narrows as the temperature is lowered.  Above 75 K the optical conductivity is frequency and temperature independent. Interestingly, we find that in the temperature range 75 K to 20 K the area under the Drude peak is {\it temperature independent} with a plasma frequency of $\approx 400$ meV. This is a signature that $m^*$ is constant in this region of temperatures. A distinct minimum develops between the Drude peak and the high frequency saturation value. We suggest this minimum is a pseudo-hybridization gap normally associated with the formation of the Kondo lattice but not fully formed in this material above 17.5 K.  There is a simple relationship between the Kondo temperature $T_K$, the effective mass $m^*$ and the gap $V_K$: $m^*/m_e=(V_K/k_B T_K)^2$ \cite{millis87,dordevic01}.  Estimating $T_K=75$ K from the temperature where the Drude peak first appears, and taking $V_K=15 \pm 5 $meV, we find  $m^*/m_e=5 \pm 2 $ which is lower than what is estimated from specific heat measurements\cite{maple86} but not in disagreement with recent STM\cite{schmidt10} or optical\cite{levallois11} data. We note here that the hybridization gap acts like the pseudogap in the cuprates.   Its frequency does not change with temperature but fills in gradually as the temperature is raised.  Also, the spectral weight lost in the gap region is not recovered by the Drude peak or in the spectral region immediately above the gap. The inset shows the accumulated spectral weight at the five temperatures.   All the curves cross at 15 meV showing that the Drude weight is conserved in the temperature range from 20 to 75 K.  On the other hand, spectral weight is lost above this frequency range as the temperature is lowered. These behaviors are inconsistent with a simple picture of an effective mass resulting from an inelastic interaction with a bosonic spectrum.

To examine quasiparticle damping above the hidden order transition we apply an extended Drude model to the conductivity:
 \begin{equation}
\sigma(T,\omega)=\frac{\omega_p^2} {4\pi} \frac{1} 
{1/\tau^{op}(\omega)-i\omega(1+\lambda(\omega))} 
\end{equation}
where $\omega_p^2=4\pi n e^2/m_e$ is the plasma frequency squared, $1/\tau^{op}(\omega)=\frac{\omega_p^2}{4\pi}{\cal R}{e}({1} / \sigma(\omega))$, the optical scattering rate, and $1+\lambda(\omega) = m^*/m_e$ is the mass enhancement. Optical phonons at 13.5 and 46.9 meV have been subtracted from the measured conductivity.  The renormalized optical scattering
rate $1/\tau^*=1/\tau^{op}/(m^*/m_e)$  is shown in figure 3 a) where we have used a plasma frequency of $\omega_p^* = \omega_p/\sqrt{m^*/m_e}=$ 418  meV, evaluated from the Drude weight.  As the temperature exceeds $T_K$, here taken as 75 K, the frequency dependence below 14 meV is replaced by uniform temperature and frequency independent scattering. We also note that the low frequency scattering above 20 K is {\it incoherent} in the sense that $1/\tau^* > \omega$ but, significantly, the condition reverses at 20 K, near the temperature of the hidden order transition.

We next turn to the optical resistivity, defined as  $\rho(\omega)={\cal R}{e}({1} / \sigma(\omega))$ where $ \sigma(\omega)$ is the complex conductivity. We used the "refined reflectivity" (see Materials and Methods) to calculate this quantity as plotted in figure 3 b) at three temperatures, but since we are focussing on temperatures just above the phase transition we use lower noise {\it refined}  data. The zero frequency limit of $\rho(\omega)$ is the dc resistivity, which, as mentioned above, has been adjusted to agree with the measured resistivity shown as filled circles at zero frequency. Figure 3 b) also shows a parabola fitted to the data where the constant $A'(T)$ and a dc offset  $c(T)$ are adjustable parameters.  

\begin{figure*}
     \vspace*{-2.5 cm}
\hspace*{-0.5cm}
\includegraphics  [width=8.7cm]{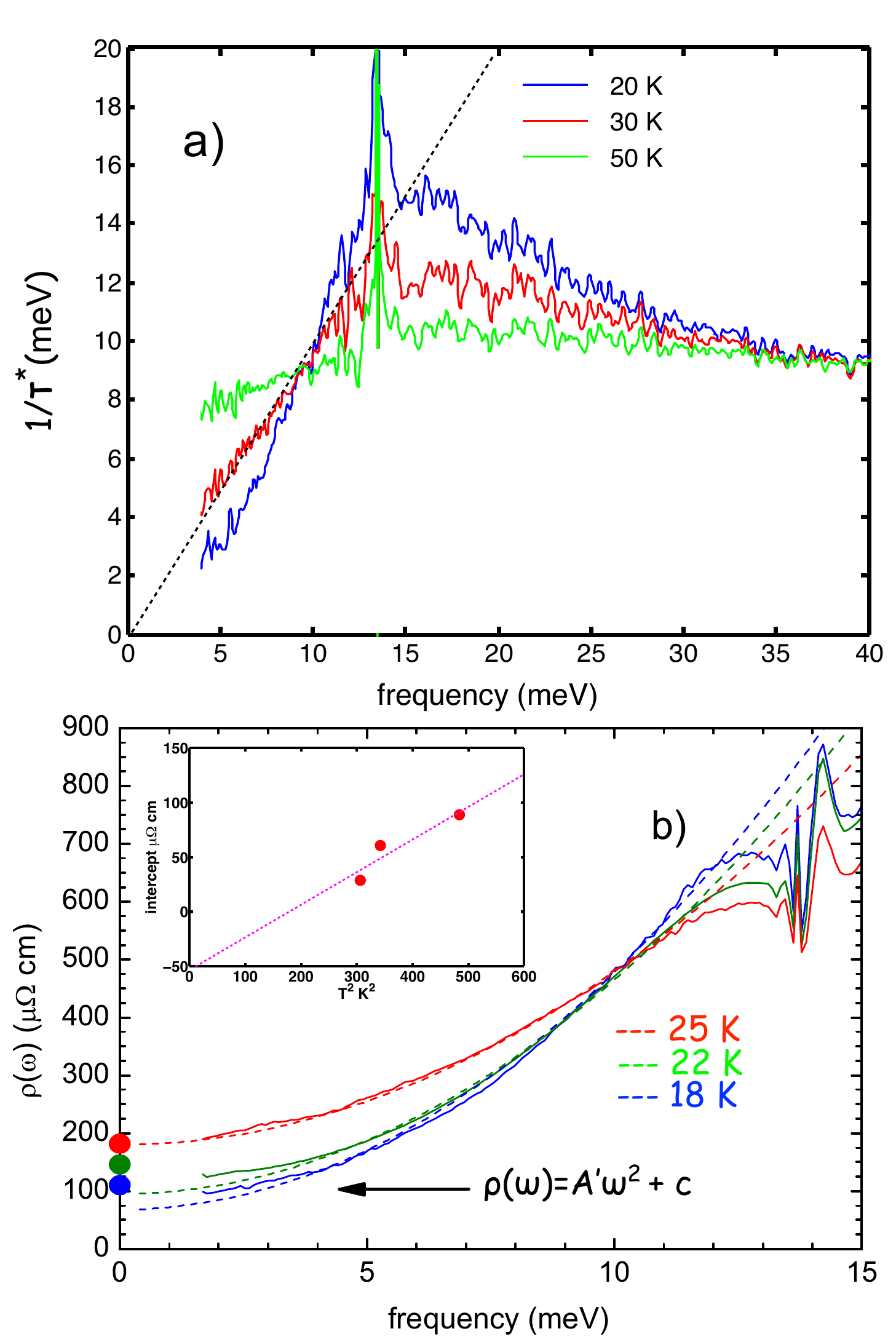}
\vspace*{-2.5 cm}%
\caption{(color online). a) The frequency dependent scattering rate $1/\tau^*$ at three temperatures in the normal state above the hidden order transition at 17.5 K from the unrefined reflectivity.  As the temperature is raised the Fermi liquid scattering  below 14 meV is replaced by a uniform frequency independent incoherent scattering.  Coherent quasiparticles exist below the dashed line $\omega > 1/\tau^*$. b) The optical  resistivity $\rho(\omega)$ vs. photon energy at low frequencies from the refined reflectivity. The experimental curves (solid lines) are compared to a Fermi liquid fit (dashed lines) with the coefficient $A'$ and an offset $c(T)$ determined by a least squares fit to the experimental data. The inset shows the temperature dependence of $c(T)$ plotted as a function of $T^2$, for the three lowest temperatures, 17.5 K 18 K and 22 K.  The slope of this curve yields an estimate of $A=0.30 \mu\Omega$ cm K$^{-2}$ from optics.}
\label{Fig3}
\vspace*{3.0cm}
\end{figure*}

\begin{figure*}
\vspace*{0.1 cm}
\hspace*{-0.5cm}
\includegraphics  [width=8.7cm]{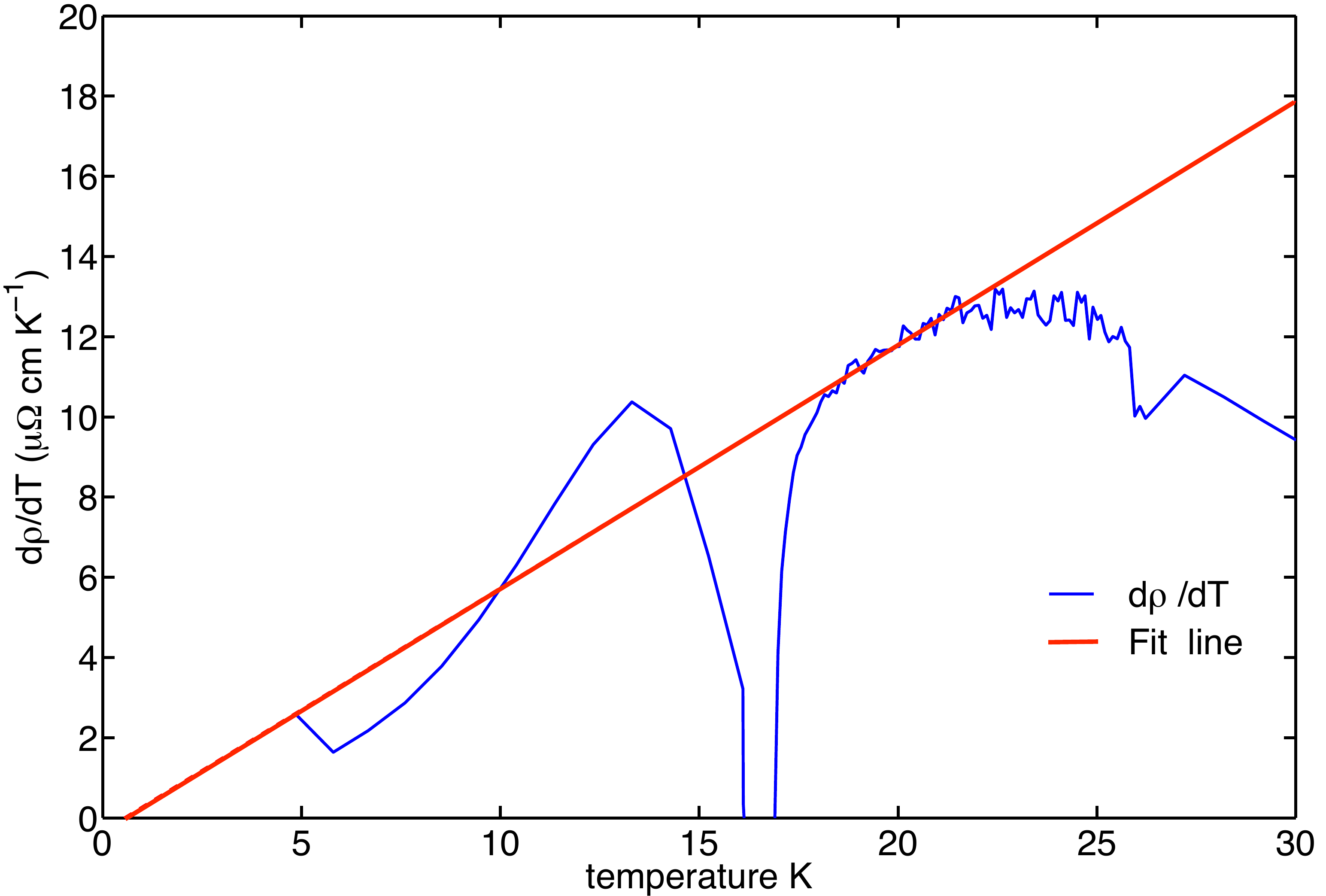}
\vspace*{0.5 cm}%
\caption{(color online).   The solid line shows the temperature derivative of the experimental dc resistivity of URu$_2$Si$_2$. The straight line is a fit of  $d\rho(T)/dT = c + 2A T$ in the temperature range 18--22 K.   Above 22 K the temperature derivative falls, a signature of the onset of incoherence.}
\label{Fig4}
\end{figure*}

We next evaluate the Fermi liquid parameters $A'$ and $A$ from our data as well as the constant $b$. We determine $A'$ directly from a quadratic fit to the optical data shown in figure 3 b) between 5 and 11 meV.    Note that the the scattering rate deviates from the simple quadratic form below 5 meV and above 12 meV where it saturates.   The coefficient $A'=0.034 \mu\Omega$ cm K$^{-2}$ at 17.5 K and decreases to 0.030 at 22 K whereas the cutoff seems to remain at 12 meV.   Even with our enhanced signal to noise ratio we see little evidence to coupling to sharp resonance modes in our spectra of the type seen in the cuprates\cite{hwang07}. Such modes, whether they are magnetic or due to phonons, would show a characteristic rise of scattering rate at the mode frequency. {\it Instead, the self energy of the quasiparticles is dominated by a featureless continuum without an energy scale}.

The inset to figure 3 b) shows the intercept $c(T)$ plotted as a function of $T^2$. The slope gives us the coefficient $A=b(\pi)^2A'=0.30 \mu\Omega$ cm K$^{-2}$ and $b=1.0 \pm 0.1$, an average over the temperature region 18.5 K to 22 K.  The intercept is negative but in view of large range of extrapolation we do not consider this significant. A positive intercept would suggest a linear $T$ contribution whereas a negative one implies a Kondo-like process that rises as the temperature is lowered.  While the scatter in the points precludes any definite conclusions it is clear from the raw data that an upward trend is present in $\rho(\omega)$ below 3 meV and below 25 K.

We next compare these optically determined parameters with the parameters determined from the dc resistivity.  Figure 4 shows the temperature derivative of the dc resistivity of URu$_2$Si$_2$. The line is a straight line fit to the derivative in the 18 to 22 K temperature range to $d\rho/dT = c + 2AT$ with $A=0.3 \pm 0.12$ $\mu\Omega$ cm K$^2$.  The fit shows that the resistivity is dominated by a $T^2$ term and the coefficient $A$ agrees with its value determined from optics well within experimental error.  The near-zero value of the intercept $c$ shows that there is only a weak linear in $T$ contribution to the scattering but it should be noted that in view of the narrow 4 K temperature range used in the fit, by itself, figure 3 does not prove that we have a Fermi liquid above 18 K.  In fact, higher resolution dc resistivity data\cite{matsuda11} shows that there is {\it no} finite region where $\rho(T)$ is linear in $T$.  If Fermi statistics and electron electron scattering dominate the resistivity and  $1/\tau < \omega$,  we expect that, in addition to the $\omega^2$ dependence of the ac and $T^2$ dependence of the dc resistivity,  the coefficient $b$ has to equal 4.  In URu$_2$Si$_2$ all the conditions are met except the last one.  Our strongest evidence for this  are the frequency fits in figure 3 b) and the main role of the dc resistivity fit is to confirm the value of the coefficient $A$. The agreement of the $A$ coefficients obtained by optics and transport is better than expected since the experiments were done on different samples from the same batch and absolute dc resistivities generally do not agree to better than 20 percent among groups.  A Fermi-liquid like resistivity above 17.5 K in URu$_2$Si$_2$ with $A=0.35$ has also been reported by Palstra {\it et al.}\cite{palstra86} Another comparison between the temperature and frequency dependence of scattering is the ratio of the Kondo temperature $T_K=75$ K and the cutoff frequency $\omega_c=14$ meV of frequency dependent scattering. If it is written as $b_c = \omega_c^2/\pi^2 T_K^2$ we find that $b_c=0.48$,  again substantially smaller than the Fermi liquid value of $b=4$.

We conclude that instead of the  expected value of $b=4$ for Fermi liquid scattering\cite{gurzhi58} our data clearly show that $b = 1 \pm 0.1$  in the temperature region immediately above the hidden order transition. This discrepancy is well outside our possible error. The value $b=1$ is expected for resonant elastic scattering from impurities\cite{Maslov}, when the single-particle scattering rate has an $\omega^2$  but no $T^2$ term. The Kubo formula then yields the optical $1/\tau$ with $b=1$. Here, however, the scattering appears to be intrinsic. One possibility is that in this material, instead of the formation of an Anderson lattice of coherent states, the uranium $f$ levels act like independent incoherent scatterers and form the coherent lattice only below the hidden order transition.  This picture has also been advanced by Haule and Kotliar\cite{haule09} and Schmidt \etal \cite{schmidt10}  Our data provide independent evidence for this model.  
The important question remains, are there cases of true Fermi liquids with $b=4$? As table 1 shows, all the cases where the frequency dependence has been measured fail to show clear cases where $b=4$. The deviation from the Fermi liquid value of $b$ has been discussed by Rosch and Howell\cite{rosch05} for some special cases, such as quasi-2D compounds and a case with  $b=5.6$, is reported by Dressel.\cite{dressel11}  
  
In summary, we have found that in the normal state above the hidden order transition in  URu$_2$Si$_2$ a relatively light band with a mass $m^*/m_e  \approx$ 5 is weakly coupled to the $f$ electrons with $V_K \approx 5$  meV, and that this band is responsible for the transport current as measured by the optical conductivity.  We suggest that this coupling is not strong enough to form an Anderson lattice. Instead the $f$ electrons act like elastic, incoherent scatterers as shown by the anomalous $b=1$ in the generalized Fermi liquid scattering formula instead of the expected $b=4$ for coherent inelastic scattering from bosonic excitations.  As suggested by the STM experiments of Schmidt \etal\cite{schmidt10},  the Fermi liquid with the heavy quasiparticles exists only below the hidden order transition. Because of the rapidly varying electronic density of states we are unable to use our technique to analyze the nature of the scattering below the hidden order transition to verify this scenario and recent transport experiments suggest a possible non-Fermi liquid behavior at low temperatures\cite{matsuda11}   We also note that this anomalous Fermi liquid behavior is shared by a number of other strongly correlated materials where magnetism appears to play a role. The possibility exists that in these systems the electron lifetime is not determined by Fermi liquid electron-electron scattering but by elastic resonant scattering and lead to the notion that a quadratic temperature dependence of the resistivity may not be a good signature of a Fermi liquid.



\begin{materials}
The single crystals of URu$_2$Si$_2$ were grown at Grenoble and at McMaster in tri-arc furnaces in an argon atmosphere.  The crystals were annealed, under UHV, at 900 C for 10 days. The ab plane cleaved surfaces were measured by standard reflectance techniques, at three separate laboratories, using an {\it in situ} gold overcoating technique\cite{homes93a}.  The absolute reflectance results of the three groups agreed to within 0.5 \%.

\begin{figure*}
\vspace*{0.0 cm}
\hspace*{-0.5cm}
\includegraphics  [width=8.7cm]{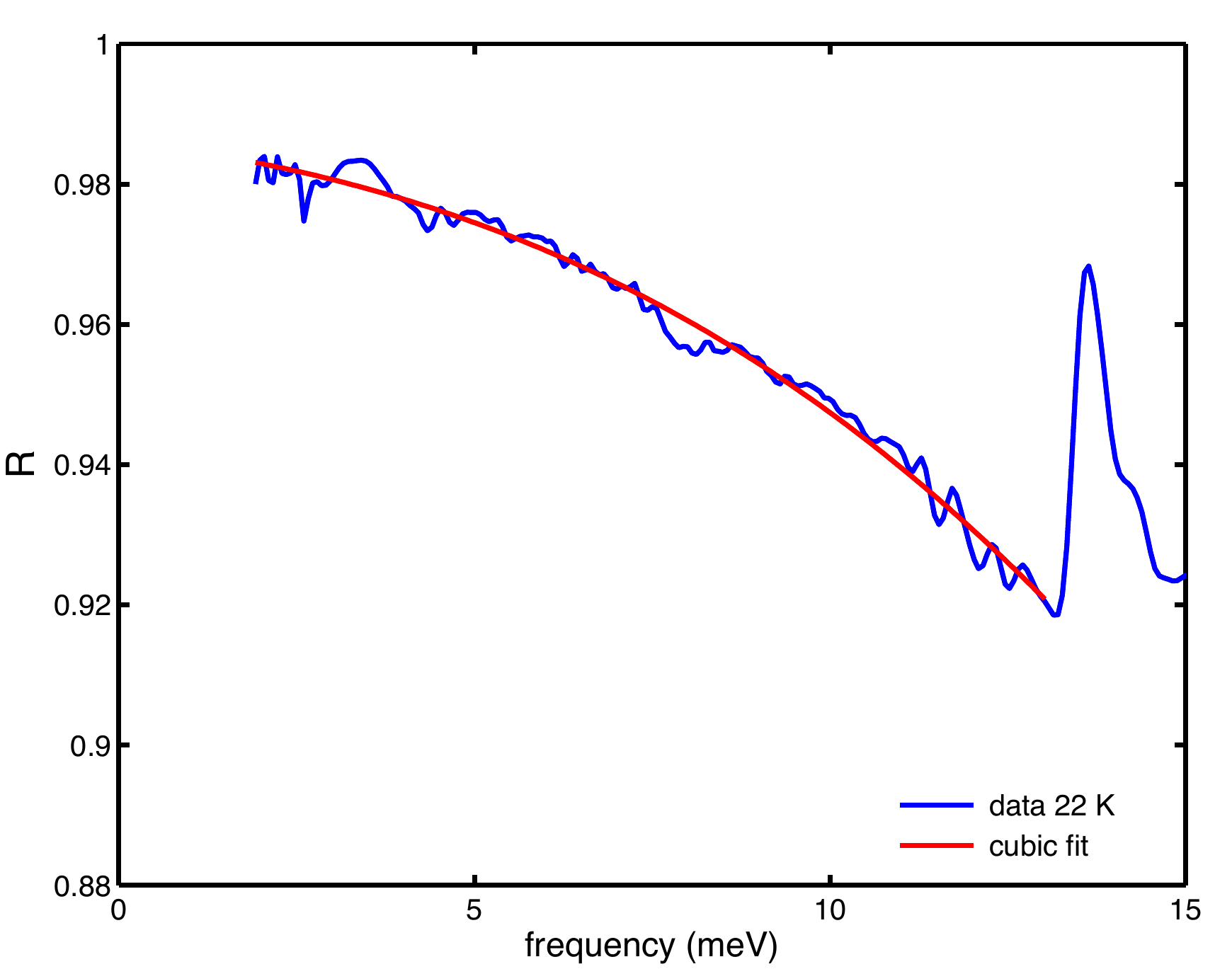}
\vspace*{0.5cm}%
\caption{(color online).  The noisy experimental reflectance data, measured at 22 K, is smoothed by fitting a cubic polynomial to the data. This curve is combined with other experimental data at higher frequencies and Kramers-Kronig transformed to yield an approximation to actual spectrum. }
\label{Fig5}
\end{figure*}

 At long wavelengths a simple procedure which we call ``refined thermal reflectance'' was used to cancel out interference artifacts\cite{purdy10} below 13 meV. The procedure involves the following steps. We have found that the interference artifacts seen in the absolute experimental spectrum, figure 5, are related to the movement of the sample stage. To overcome this we measure the reflected spectra over a narrow temperature range without moving the sample stage, typically from 4 K to 25 K.  Using one of the spectra as a reference we record the ratios of the spectra at the various temperatures to the spectrum at the reference temperature. To obtain a low noise absolute reflectance we use the gold overcoating technique to get an estimate of the absolute reflectance at the reference temperature.  Because the sample is moved in this process this absolute spectrum is contaminated by interference artifacts. To eliminate these we fit the absolute reflectance at the reference temperature with a cubic polynomial to produce a smooth curve, averaging out the artifacts, figure 5, curve labelled 'cubic fits'. This smoothed spectrum is then used as a reference spectrum to calculate absolute spectra at all other temperatures. It is clear that the smoothing procedure hides any sharp structure in the reference spectrum.  However any new sharp structure that appears as the temperature is changed will be present at full resolution.  The final refined spectra are shown in figure 6. This procedure is well suited to the discovery of new spectral features that appear at phase transitions, for example the prominent minimum at 5 meV due to the hidden order gap.  The measured refined reflectance was converted to an optical conductivity by Kramers-Kronig analysis.  At low frequency, below 4 meV,  a Drude  response was assumed where  we used the measured dc resistivity to determine the amplitude of the Drude peak and the absorption at our lowest measured infrared frequency to determine the width.  At high frequency,  beyond 7 eV, we used the results of Degiorgi {\em et al.} \cite{degiorgi97}.  

\begin{figure*}
\vspace*{0.5 cm}
\hspace*{-0.5cm}
\includegraphics  [width=8.7cm]{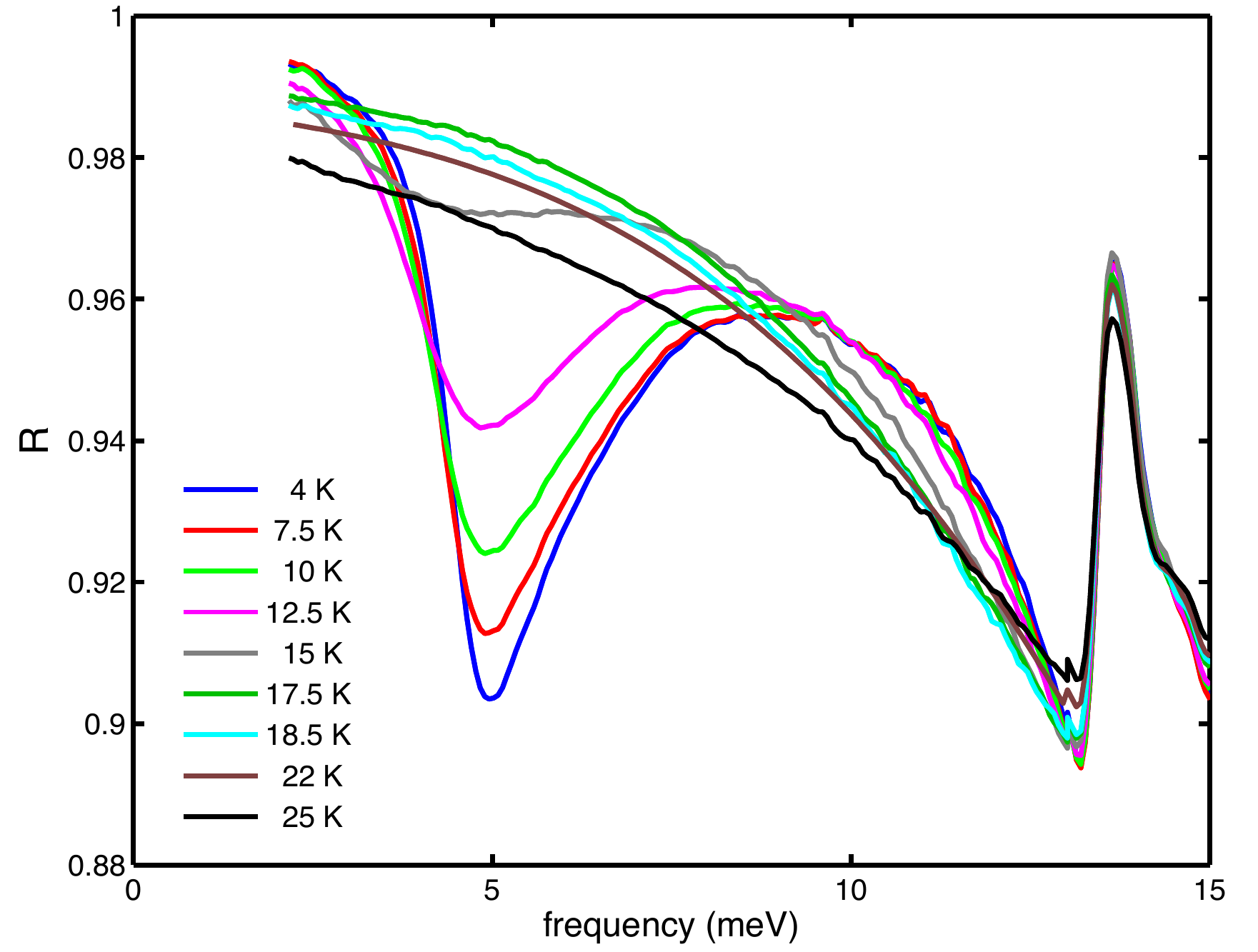}

\vspace*{-1.5 cm}%
\caption{(color online).  "Refined" reflectance of  URu$_2$Si$_2$ obtained by first measuring a series of spectra at different temperatures $T$ and then dividing the spectra with one measured at a reference temperature $T_{ref}$.  All this is done without moving the sample stage. The resulting temperature ratios are smooth without interference artifacts.  Then these smooth ratios are multiplied by the estimate to the absolute spectrum at $T_{ref}$ shown in figure 4. The resulting spectra shown in the figure are a low noise  approximation to the true absolute spectra of URu$_2$Si$_2$. Of all the spectra shown, only the one at $T_{ref}=22$ K is a polynomial fit, all the others show actual measured data.}
\vspace*{1.5 cm}
\label{Fig6}
\end{figure*}

\end{materials}


\begin{acknowledgments}
We thank  K. Behnia, D.A. Bonn, S. Bulaevsky, J.C. Carbotte, A.V.  Chubukov, P Coleman, J.C. Davis, B. Gaulin,  B. Maple, D.L.  Maslov,  A.J. Millis and D.W.L. Sprung for helpful discussions. In particular we thank K. Behnia and T. Matsuda for supplying us with unpublished data. This work has been supported by the Natural Science and Engineering Research Council of Canada and the Canadian Institute for Advanced Research. Work in Tallinn was supported by the Estonian Ministry of Education and Research under Grant SF0690029s09, and Estonian Science Foundation under Grants  ETF8170 and ETF8703 and in Paris by the ANR under Grant No. BLAN07-1-183876 GAPSUPRA.

\end{acknowledgments}






\end{article}









\end{document}